\documentclass[%
twocolumn,
 amsmath,amssymb,
 aps,
 epstopdf,
 pdflatex,
]{revtex4-1}

\usepackage{graphicx}% Include figure files
\usepackage{dcolumn}% Align table columns on decimal point
\usepackage{bm}% bold math
\usepackage{bbm}
\usepackage{siunitx}
\usepackage{mathtools}
\usepackage{braket}
\usepackage{chemformula}
\usepackage{tabls}
\usepackage[toc,page]{appendix}
\usepackage{verbatim}

\newcommand{\Rcal}{\mathcal{R}}

\newcommand{\bE}{\boldsymbol{E}}
\newcommand{\bmu}{\boldsymbol{\mu}}

\newcommand{\bZ}{\boldsymbol{Z}}

\newcommand{\bR}{\boldsymbol{R}}

\newcommand{\bD}{\boldsymbol{D}}

\newcommand{\bG}{\boldsymbol{G}}

\newcommand{\bu}{\boldsymbol{u}}

\newcommand{\bq}{\boldsymbol{q}}
\newcommand{\bk}{\boldsymbol{k}}

\newcommand{\bRcal}{\boldsymbol{\Rcal}}

\newcommand{\epsel}{\epsilon}
\newcommand{\bepsel}{\boldsymbol{\epsilon}}
\newcommand{\br}{{\bm r}}

\newcommand{\ba}{{\boldsymbol{a}}}
\newcommand{\bb}{{\boldsymbol{b}}}
\newcommand{\bc}{{\boldsymbol{c}}}

\newcommand{\eqname}{{Eq.}}

\newcommand{\bpm}{\begin{pmatrix}}
\newcommand{\epm}{\end{pmatrix}}

\newcommand{\dielectric}{\bepsel}
\newcommand{\ldielectric}{\epsel}

\renewcommand{\figurename}{{Fig.}}

\DeclareMathOperator{\erf}{erf}
\DeclareSIUnit\rydberg{\text{Ry}}

\begin{document}

	\preprint{APS/123-QED}
	
	\title{Electrostatic interactions in atomistic and machine-learned potentials for polar materials}% Force line breaks with \\
	
	\author{Lorenzo Monacelli}
	\author{Nicola Marzari}
	%\email{lorenzo92monacelli@gmail.com}
	
	%\author{Antonio Siciliano}
	
	%\author{Francesco Mauri}%
	\affiliation{Theory and Simulation of Materials (THEOS), and National Centre for Computational Design and Discovery of Novel Materials (MARVEL), École Polytechnique Fédérale de Lausanne, 1015 Lausanne, Switzerland}%Lines break automatically or can be forced with \\

	\date{\today}% It is always \today, today,

\begin{abstract}
    Long-range electrostatic interactions critically affect polar materials. However, state-of-the-art atomistic potentials, such as neural networks or Gaussian approximation potentials employed in large-scale simulations, often neglect the role of these long-range electrostatic interactions. 
    This study introduces a novel model derived from first principles to evaluate the contribution of long-range electrostatic interactions to total energies, forces, and stresses. The model is designed to integrate seamlessly with existing short-range force fields without further first-principles calculations or retraining. The approach relies solely on physical observables, like the dielectric tensor and Born effective charges, that can be consistently calculated from first principles. We demonstrate that the model reproduces critical features, such as the LO-TO splitting and the long-wavelength phonon dispersions of polar materials, with benchmark results on the cubic phase of barium titanate (\ch{BaTiO3}).
\end{abstract}
	
	\maketitle
    \section{Introduction}
In polar materials, displacing an ion from equilibrium generates an electric dipole. The resulting electric field decays as the inverse of the cube of the distance and interacts with local dipoles generated by far-away ionic displacements.
While these long-range electrostatic interactions are screened in metals, and their role becomes relevant only when accounting for ionic dynamics\cite{binci_first-principles_2021,marchese_born_2024}, in the absence of electrons in the conduction band long-range electric fields survive down to zero frequency, thus playing a fundamental role in the phenomenology of insulators. This includes the splitting between longitudinal and transverse optical phonons at long wavelengths (LO-TO splitting)\cite{cochran_dielectric_1962,giannozziInitioCalculationPhonon1991, gonze_interatomic_1994,gonze_dynamical_1997}, affecting in turn all thermodynamic properties related to phonon dispersions, such as thermal expansion, heat capacity, lattice thermal conductivity, and Raman and IR spectra.
First-principles simulations include long-range electrostatic interactions~\cite{royoExactLongRangeDielectric2021,giannozziInitioCalculationPhonon1991} when considering perturbations of finite wavevector, where the phonons can be evaluated either by finite-differences\cite{phonopy} or density-functional perturbation theory (DFPT)\cite{giannozziInitioCalculationPhonon1991}. At $\Gamma$ ($\bq=0$), the contribution of electrostatic interaction is nonanalytic, as $\bq=0$ periodic displacements produce macroscopic polarizations resulting in a dynamical matrix for phonons dependent on the direction of the $\bm q$ vector as it approaches $\Gamma$.
This nonanalytic behavior of lattice force constants near $\Gamma$ constitutes a challenge in modeling efficient force fields for polar materials, as it originates from the cubic power-law decay of electrostatic forces and can not be captured if the interaction is truncated over a finite distance.
%To Fourier interpolate the nonanalytic lattice force constants near $\Gamma$, a model for electrostatic dipole interactions is employed to remove the long-range part in the finite $\bq$ force constants. In this way, the force-constant matrix is divided into a short-range part that behaves analytically and can be easily Fourier interpolated and a nonanalytical long-range one which can be computed at any $\bq$ point\cite{Baroni2001}. 
%However, this is impossible in models parametrizing the atomistic potential fitted on first-principles data on periodic boundary conditions (PBC), as they lack the long-range contribution of forces, which are relevant only for q points near $\Gamma$, thus commensurate with large supercell. Thus, besides the need for a proper parametrization that may take into account long-range interactions, atomistic models must be fitted on first-principle data with PBC on sufficiently big cells.

Thanks to the advances in machine learning technologies, many excellent tools to fit atomistic potentials from small sets of high quality \emph{ab initio} calculations have emerged\cite{Behler2016,schutt_schnet_2017, wang_deepmd-kit_2018,deringer_gaussian_2021,kulik_roadmap_2022,batzner_e3-equivariant_2022}, paving the way to the simulation of large-scale materials with hundreds of millions of atoms with an accuracy close to first principles methods\cite{Lu2021,allegro}.
However, the most commonly employed machine-learning potentials neglect the long-range electrostatic interactions, resulting in a smooth $\bm q$ dependence of the atomic force constant matrix around $\Gamma$. This is because the force field parametrizations account only for local atomic environments, and the first principles training data are evaluated on periodic boundary conditions that lack long-range contributions. The question arises of what should be fitted and how electrostatics can be recovered. Several methods have been developed to overcome this issue. Some approaches, the so-called second and third-generation machine-learning potential\cite{Behler2021}, split the energy expression into a short and long-range part, explicitly accounting for electrostatic interactions between point charges dependent on the local nuclear environment\cite{Artrith2012,Sifain2018,Ghasemi2015, Bleiziffer2018,Nebgen2018,Yao2018}. These models suffer from the need for extra training data like partial atomic charges that, while easy to extract from DFT simulations, are not physical observables and ubiquitously defined\cite{Sifain2018}. To overcome this issue, different methods have been proposed to rely on physical observables, like the electron densities\cite{Chandrasekaran2019}, the Wannier functions\cite{Zhang2022}, and polarization\cite{Gastegger2017,Grisafi2021}, which can be rigorously defined through the modern theory of polarization\cite{KingSmith1993,Resta1994,RestaVanderbilt}. These approaches have also been augmented to describe charge relocalization by solving a charge equilibration equation self-consistently, the so-called 4th generation scheme\cite{ko_fourth-generation_2021,shaidu_incorporating_2024}. While promising, these procedures require new first-principles calculations to store the extra information needed for the training, hampering the reuse of energy/forces data generated, for example, from previous first-principles calculations of molecular dynamics (FPMD). 
A different strategy introduces long-range descriptors of atomic coordinates\cite{Grisafi2019}, also achieved by message-passing neural architectures architectures\cite{Zubatyuk2021}. While very promising, especially in their ability to account for macroscopic rearrangement of the electron charge in the system, these models need a training set with sufficiently large simulation cells to sample long-range interactions\cite{Grisafi2019,Grisafi2021,Zubatyuk2021}, preventing the employment of small simulation cells in the training data.
  %Moreover, the solutions available often cannot be combined with different short-range force fields.

This work presents instead a comprehensive first-principles approach designed to integrate long-range interactions seamlessly with existing short-range force fields in polar condensed matter systems. The method accurately calculates total energies, forces, and stress tensors, and it is grounded solely on a select set of well-established first-principles properties of the material in its equilibrium state.

%In \secname~\ref{sec:electrostatic:energy}, we introduce the theoretical framework and derive the expression for the electrostatic energy. The forces and stress tensors are presented in \secname~\ref{sec:force} and \ref{sec:stress}, respectively. Then, we discuss the long-wavelength limit and how the derived forces correctly reproduce the LO-TO splitting. We report a calculation of the phonon bands in the cubic phase of \ch{BaTiO3} comparing a state-of-the-art GAP potential\cite{gigliThermodynamicsDielectricResponse2022} with and without the addition of the long-range forces with the \emph{ab initio} reference, showing how our approach efficiently fixes the phonon dispersion near $\Gamma$.
    
\section{Results}%The electrostatic energy}
\subsection{Electrostatic energy}
\label{sec:electrostatic:energy}
The coupling between dipoles is the leading term of the long-range electrostatic interactions in neutral systems. Each dipole $\bmu_i$ generates an electric field $\bE(\bR_i)$ decaying as the inverse of the cube of the distance, which interacts with other dipoles as
\begin{equation}
\mathcal{E} = -\frac 12\sum_{i=1}^N \bE(\bR_i) \cdot\bmu_i,
\label{eq:energy}
\end{equation}
where bold symbols represent vectors and matrices; the $\frac 12$ in \eqname~\eqref{eq:energy} prevents double counting when summing over all the atoms. 

The overall dipole moment $\bmu$ of the crystal can be rigorously defined through the modern theory of polarization\cite{KingSmith1993,Resta1994,RestaVanderbilt}. Here, we can define the atomic dipole $\bmu_i$ as the contribution to the overall dipole moment $\bmu$ given by the displacement of atom $i$
\begin{equation}
Z_{i\alpha\beta} = \frac{\partial \mu_\alpha}{\partial R_{i\beta}},
\qquad 
\mu_\alpha(\bR) = \mu^{(0)}_\alpha + \sum_{i=1}^N \mu_{i\alpha}(\bR),
\end{equation}
\begin{equation}
    \mu_{i\alpha}(\bR) = \sum_\beta Z_{i\alpha\beta} (R_{i\beta} - \Rcal_{i\beta})
    \label{eq:dipole:moment},
\end{equation}
where $\mu_\alpha$ and $\mu_{i\alpha}$ are the $\alpha$ Cartesian component of the dipole moment per cell of the total system and the one of the $i$-th atom, respectively. $Z_{i\alpha\beta}$ is the Born effective charges tensor, $\boldsymbol{\Rcal_{i}}$ is the equilibrium position of the $i$-th atom, and $\bm \mu^{(0)}$ is the total dipole moment when atoms are in the equilibrium position, defined up to a quantum of polarization. We identify the atoms with Latin indices and the Cartesian components with Greek ones.
%Here we use Latin letters for atomic indices ad greek letters for Cartesian components.
%We assume the net dipole moment in the equilibrium configuration equals zero. This is correct in the long-range limit of the atomic forces, as the contribution of the microscopic electric field variation due to far-away dipoles decay quickly with the distance between atoms.
%Substituting the expression of the dipoles (\eqname~\ref{eq:dipole:moment}) in the energy per cell $\mathcal{E}$ we get
%\begin{equation}
%\mathcal{E}= -\frac 12\sum_{i\alpha\beta} (R_{i\alpha} - \Rcal_{i\alpha}) Z_{i\beta\alpha} E_{\beta}\left(\bR_i\right).
%\label{eq:energy}
%\end{equation}
%where $E_\beta$ is the $\beta$ Cartesian component of the electric field on the atom. We employ a point-charge representation of the dipole to evaluate the electric field.
To parametrize the electric field generated by the dipoles of \eqname~\eqref{eq:dipole:moment}, we define an auxiliary distribution of spherical charges reproducing a dipole $\bmu_i$ on each atom. In particular, each dipole consist of two charges $q$ of opposite signs and at a distance $d$ as
\begin{equation}
\label{eq:mu:simple}
\bmu = q \boldsymbol{d} \;.
\end{equation}
The modulus of the dipole $qd$ leaves an arbitrary choice on the values for $q$ and $d$, which, however, only affects terms of the multipole expansions beyond the dipole and, thus, disappears from the final expression for the energy and its derivatives (forces and stresses) in the dipole limit (\appendixname~\ref{app:energy}). %Since we are considering only the electric field generated from far-away dipoles, the effect of nearby charges operates as a dielectric medium partially screening the interaction. 
%
%\begin{equation}
%\bD(\br) = \frac{q}{4\pi} \frac{\br - \tilde\bR}{\left|\br- \tilde\bR\right|^3}  
%\end{equation}
%We are interested in the electric field, which is obtained as:
%\begin{equation}
%\bD= \varepsilon_0 \dielectric\bE
%\end{equation}
%Here $\dielectric$ is the infinite frequency dielectric constant: the dielectric constant only due to the electronic contribution, neglecting the one arising from atomic displacements. It could be a tensor, which means that $\bD$ and $\bE$ are not necessarily parallel.
\begin{comment}
The microscopic electric field is the sum of the electric field generated by all the point charges in the auxiliary electrostatic system:
\begin{equation}
\bE(\br) = \sum_i\frac{q_i}{4\pi\varepsilon_0}\frac{\dielectric^{-1} \left(\br -\tilde\bR_i\right)}{\left|\br- \tilde\bR_i\right|^3} 
\label{eq:electric:field}
\end{equation}
where $\dielectric$ is the static dielectric tensor (also obtained with a single linear response calculation on the average configuration).
\end{comment}

By exploiting this system of charges, we define a charge density $\rho(\bm r)$ that reproduces the correct long-range dipole-dipole interactions but sufficiently slow-varying to smear out short-range interactions.
\begin{equation}
\rho{\left(\bm r \right)} = \sum_{j = 1}^{2N} \frac{ q_j }{\sqrt{8 \pi^3} \eta^{2}}\exp\left[- \frac{(\br - \tilde\bR_j)^{2}}{2 \eta^{2}}\right],
\label{eq:rho:r}
\end{equation}
where $\tilde\bR_j$ is the position of the charge $q_j$. We derive an expression for $\tilde\bR_j$ in \appendixname~\ref{app:charge:system}.
The $\eta$ parameter is the short-range smearing. This way, the resulting force field does not affect energies and its derivatives computed on periodic cells with a linear size smaller than $\eta$. 
This approach is highly practical as it enables adjusting $\eta$ in such a way that the charge density electrostatic interactions do not affect training data defined in small cells employed to develop the short-range machine-learning interatomic potential. Consequently, this makes the potential reusable in conjunction with our charge model, eliminating the need for retraining.

%\begin{equation}
%4\pi\int_0^\infty {r}^2 dr \rho(r) = q.
%\end{equation}

\begin{comment}
The electric field generated by this charge distribution is
\begin{align}
\bE(\br) = \sum_i \frac{q_i}{4\pi}&\frac{\dielectric^{-1}\tilde\br_i}{\left|\tilde\br_i\right|^3} \left[
\erf{\left(\frac{\sqrt{2} \tilde r_i}{2 \eta} \right)} - \frac{\sqrt{2} \tilde r_i e^{- \frac{{\tilde r_i}^{2}}{2 \eta^{2}}}}{\sqrt{\pi} \eta}
\right] \nonumber\\
\tilde \br_i = \br - \tilde{\bR_i}&
\label{eq:total:electric:field}
\end{align}
where $\dielectric$ is the static high-frequency dielectric tensor (including $\varepsilon_0$). \eqname~\eqref{eq:total:electric:field} recover the correct long-range electric field for $|\br - \tilde\bR_i|\gg \eta$.
\end{comment}

The electric field produced by a polar charge distribution, as in \eqname~\eqref{eq:rho:r}, is slowly and conditionally convergent as it generates a macroscopic charge on the surface of the solid. The problem is solved in Fourier space with the Ewald summation for a 3D bulk material (details in \appendixname~\ref{app:efourier}):
\begin{equation}
    \bE(\br) = \frac{i}{\Omega}\sum_{\substack{j\\ k_j\neq 0}} \frac{\bk_j e^{-\frac{\eta^2k_j^2}{2}}e^{i\bk_j\cdot \br}}{\sum_{\alpha\beta}{k_j}_\alpha \ldielectric_{\alpha\beta} {k_j}_\beta}S(\bk_j),
    \label{eq:E:fourier}
\end{equation}
where $\Omega$ is the supercell volume, $i$ is the imaginary unit coming from the Maxwell-Equations in Fourier space, and $S(\bk)$ is the structure factor associated to the auxiliary charge system:
\begin{equation}
    S(\bk) = \sum_{j=1}^{2N} q_j e^{-i\bk\cdot \tilde\bR_j}.
\label{eq:Sk}
\end{equation}
The values assumed by the $\bk$ vectors are constrained to be multiples of the reciprocal lattice:
\begin{equation}
\bk_{(l, m, n)} = l\ba + m \bb + n\bc, \label{eq:k:vals}
\end{equation}
where $\ba$, $\bb$, and $\bc$ are the reciprocal lattice vectors of the periodic supercell, and $l,m,n$ go from $-\infty$ to $\infty$, excluding $\bk = 0$. Additional care must be taken if the charge distribution is not defined in bulk systems, as in 2D materials or 1D chains, where \eqname~\eqref{eq:E:fourier} is no longer valid and a nonuniform spatial dependency of the dielectric tensor needs to be taken into account\cite{Sohier2017,royoExactLongRangeDielectric2021,Rivano2023}.

Integrating the electrostatic energy of \eqname~\eqref{eq:rho:r} with the expression of the electric field (\eqname~\ref{eq:E:fourier}), we get the total energy (see \appendixname~\ref{app:energy} for details on the calculation)
\begin{align}
    \mathcal E(\bm R) = \frac 12\sum_{ij\alpha\beta\mu\nu}&(R_{i\alpha} - \Rcal_{i\alpha})(R_{j\mu} - \Rcal_{j\mu})\frac{Z_{i\beta\alpha}Z_{j\nu\mu}}{\Omega} \cdot \nonumber \\ &\cdot\sum_{\substack{k\\k\neq 0}}\frac{k_\beta k_\nu e^{-\frac{\eta^2k^2}{2}}}{\sum_{\mu\nu} k_\mu\ldielectric_{\mu\nu}k_\nu} e^{-i\bk(\bR_j - \bR_i)} 
    \label{eq:energy:final}\; .
\end{align} 

\eqname~\eqref{eq:energy:final} is the central result of this work. It defines the energy contribution of long-range dipole-dipole interactions and depends explicitly on the atomic positions. Notably, the expression for the energy only depends on physical observables, as the arbitrary (nonphysical) values of the atomic charges $q_j$ and their positions $\tilde\bR_j$ cancel out in the dipole limit.
The computation of \eqname~\eqref{eq:energy:final} requires first-principles quantities like the high-frequency dielectric tensor $\dielectric$ and effective charges $\bZ$ at the equilibrium position $\bm \Rcal$, and the only free parameter is the smearing factor $\eta$. 

% The overall charge within r
%\begin{align}
%Q(r) &= 4\pi\int_0^r {r'}^2 dr' \rho(r') \nonumber\\
%&=  q \left[\operatorname{erf}{\left(\frac{\sqrt{2} r}{2 \eta} \right)} - \frac{\sqrt{2} r e^{- \frac{r^{2}}{2 \eta^{2}}}}{\sqrt{\pi} \eta}\right]
%\label{eq:q:r}
%\end{align}

    %\input{src/polarizability.tex}
    
\subsection{Forces and stress tensor}
\label{sec:force}
Long-range forces are obtained by deriving the expression for the electrostatic energy:
\begin{equation}
    f_{i\alpha} = -\frac{\partial \mathcal E}{\partial R_{i\alpha}}.
\end{equation}
\begin{align}
f_{i\alpha} &= -\sum_{j\beta\mu\nu}(R_{j\mu} - \Rcal_{j\mu})\frac{Z_{i\beta\alpha}Z_{j\nu\mu}}{\Omega} \cdot \nonumber \\ &\qquad\qquad \cdot \sum_{\substack{k\\k\neq 0}}\frac{k_\beta k_\nu e^{-\frac{\eta^2k^2}{2}}}{\sum_{\mu\nu} k_\mu\ldielectric_{\mu\nu}k_\nu} \cos\left[\bk(\bR_j - \bR_i)\right] + \nonumber \\
&\sum_{j\beta\gamma\mu\nu}(R_{i\gamma} - \Rcal_{i\gamma})(R_{j\mu} - \Rcal_{j\mu})\frac{Z_{i\beta\gamma}Z_{j\nu\mu}}{\Omega} \cdot \nonumber \\ &\qquad\qquad\cdot\sum_{\substack{k\\k\neq 0}}\frac{k_\alpha k_\beta k_\nu e^{-\frac{\eta^2k^2}{2}}}{\sum_{\mu\nu} k_\mu\ldielectric_{\mu\nu}k_\nu}\sin\left[\bk(\bR_j - \bR_i)\right]
\label{eq:forces}
\end{align}
The only term that survives in the long-wavelength limit is the first one, as the second one goes as $k^2$ for small values of $k$. However, in the actual implementation, we kept the full expression \eqname~\eqref{eq:forces} to guarantee that the numerical value of the forces coincides with the gradient of the total energy. The partition of the total dipole moment of the cell into local atomic dipoles operated in \eqname~\eqref{eq:dipole:moment} violates the translational invariance of the system, introducing a nonzero net force in the center of mass.
However, the overall translational invariance can be easily restored directly in the total energy (\eqname~\ref{eq:energy:final}) by redefining the centroids $\bRcal$ as a function of the atomic positions to eliminate any rigid translation from the displacement $\bR - \bRcal$ (see \eqname~\ref{eq:rcal:asr}).
In \appendixname~\ref{app:asr}, we show how this choice restores the full translational invariance, correcting forces and stresses to satisfy the translational acoustic sum rule.

\label{sec:stress}
The stress tensor $\bm \sigma$ quantifies the energy to deform the lattice. Its computation is required in variable cell simulations, like NPT molecular dynamics or in the stochastic self-consistent harmonic approximation (SSCHA)\cite{monacelliPressureStressTensor2018, monacelliStochasticSelfconsistentHarmonic2021,monacelli_simulating_2024,siciliano_beyond_2024,miotto_fast_2024}, and is fundamental in evaluating thermal expansion and the equation of state; it is defined
\begin{equation}
    \label{eq:stress:original}
    \sigma_{\alpha\beta} = 
    - \frac{1}{\Omega} \frac{\partial \mathcal E}{\partial \varepsilon_{\alpha\beta}},
\end{equation}
where the strain tensor $\bm\varepsilon$ is a symmetric matrix. Atomic positions (clamped ions) follow strain as
\begin{equation}
    R_\alpha'(\bm \varepsilon) = R_\alpha + \sum_\beta\varepsilon_{\alpha\beta}R_\beta.
    \label{eq:R:strain}
\end{equation}
Since strain changes the lattice parameters, also the reciprocal vectors $\bm k$ are affected as
\begin{equation}
    k'_\alpha(\bm \varepsilon) = k_\alpha - \sum_\beta \varepsilon_{\alpha\beta}k_\beta ;
    \label{eq:k:strain}
\end{equation}
therefore, the scalar product between $\bR$ and $\bk$ remains unchanged under strain. The volume is also affected by the strain as
\begin{equation}
    \Omega'(\bm \varepsilon) = \Omega\left(1 + \sum_{\alpha= 1, 3}\varepsilon_{\alpha\alpha}\right).
\end{equation}
The expression of the stress can be easily obtained by substituting $\Omega'(\varepsilon)$, $\bR'(\varepsilon)$, $\bRcal'(\varepsilon)$, and $\bk'(\varepsilon)$ in the electrostatic energy (\eqname~\ref{eq:energy:final}), and evaluating the Jacobian of \eqname~\eqref{eq:stress:original} by applying the differential chain rule recursively.
Our implementation achieved the final result by exploiting the algorithmic differentiation as implemented in the Julia library ForwardDiff.jl\cite{ForwardDiff}.%, reported in \eqname~\eqref{eq:stress}. More 
%details on the calculation can be found in \appendixname~\ref{app:stress}.

    \subsection{LO-TO splitting}

One of the essential features that the long-range electrostatic force field must reproduce is the correct phonon dispersion in the long-wavelength limit, where, in bulk 3D material, the cubic decay of the dipole-dipole forces with the inverse of the distance gives rise to a discontinuity of the dynamical matrix at $\Gamma$, where the $\bq\rightarrow 0$ limit is dependent on the direction $\bq/|\bq|$, and consequently to the LO-TO splitting (see ref.\cite{sohier_breakdown_2017} and ref.\cite{Rivano2023} for the 2D and 1D case, respectively). 
To prove that the model correctly reproduces the phonon dispersions in the $\bm q\rightarrow 0$ limit, we derived the expression of the interatomic force constant matrix
\begin{equation}
\Phi_{\alpha\beta}^{ij} =\frac{d^2\mathcal E}{dR_{i\alpha}dR_{j\beta}} = -\frac{df_{i\alpha}}{dR_{j\beta}}.
\end{equation}
Restricting ourselves to the equilibrium position $\bR = \bRcal$, the only surviving term is:
\begin{align}
\Phi_{\alpha\beta}^{ij} & = 
\frac{1}{\Omega}\sum_{\substack{k\mu\nu\\ k_j\neq 0}} \frac{{k}_\nu{k}_\mu Z_{j\nu\beta}Z_{i\mu\alpha}e^{-\frac{\eta^2k^2}{2}}}{\sum_{\alpha\beta}{k}_\alpha \ldielectric_{\alpha\beta} {k}_\beta}\cos[\bk(\bR_i - \bR_j)] %- \nonumber \\% e^{-i \bk\cdot(\bR_i - \bR_j)} 
%& \frac 2 \Omega \sum_{\substack{k\mu\nu t\gamma\\ k\neq 0}} \frac{{k}_\nu{k}_\mu k_\beta Z_{t\nu\gamma}Z_{i\mu\alpha}e^{-\frac{\eta^2k^2}{2}}}{\sum_{\alpha\beta}{k}_\alpha \ldielectric_{\alpha\beta} {k}_\beta}(R_{t\gamma} - \Rcal_{t\gamma})(\delta_{ti} - \delta_{jt})\sin[\bk(\bR_i - \bR_j)]
\end{align}
To highlight the long-wavelength limit, we perform the Fourier transform of $\Phi_{\alpha\beta}^{ij}(\bR_i, \bR_j)$
\begin{align}
\bD_{\alpha\beta}^{ij}(\bq) =&
 \frac{1}{2\Omega}\sum_{\substack{k\mu\nu\\ \bk = \bq + \bG}} \frac{{k}_\nu{k}_\mu Z_{j\nu\beta}Z_{i\mu\alpha}e^{-\frac{\eta^2k^2}{2}}}{\sum_{\alpha\beta}{k}_\alpha \ldielectric_{\alpha\beta} {k}_\beta}e^{ik(R_i - R_j)} + \nonumber \\ 
 & \frac{1}{2\Omega}\sum_{\substack{k\mu\nu\\ \bk = -\bq + \bG}} \frac{{k}_\nu{k}_\mu Z_{j\nu\beta}Z_{i\mu\alpha}e^{-\frac{\eta^2k^2}{2}}}{\sum_{\alpha\beta}{k}_\alpha \ldielectric_{\alpha\beta} {k}_\beta}e^{-ik(R_i - R_j)}.
 \label{eq:dynamical:matrix}
\end{align}
This equation is similar to the \emph{ansatz} employed to perform the Fourier interpolation of phonons in first-principles codes like Quantum ESPRESSO\cite{giannozziInitioCalculationPhonon1991,giannozziQUANTUMESPRESSOModular2009,Giannozzi2017} or ABINIT\cite{gonze_interatomic_1994,gonze_dynamical_1997,abinit} and reproduces the correct $\bq\rightarrow 0$ limit when $|\bG| \gg \frac 1 \eta$:
\begin{equation}
\lim_{\bq\rightarrow 0}
D_{\alpha\beta}^{ij}(\bq) =
\frac{1}{\Omega} \frac{\sum_{\mu\nu} Z_{j\nu\beta}{q}_\nu{q}_\mu Z_{i\mu\alpha}}{\sum_{\mu\nu}{q}_\mu \ldielectric_{\mu\nu} {q}_\nu},
\label{eq:LOTO}
\end{equation}
which coincides with the standard expression implemented in common electronic-structure packages for the non-analytic part of the force constant matrix (\eqname~18 of Ref.~\cite{giannozziInitioCalculationPhonon1991}), after the conversion in CGS units where $\epsilon_0 = (4\pi)^{-1}$. 
\eqname~\eqref{eq:LOTO} demonstrates that the forces (\eqname~\ref{eq:forces}) and energies (\eqname~\ref{eq:energy:final}) parametrized by this model correctly reproduce the LO-TO splitting of any materials.

\subsection{Fitting the model}

The model can be employed on top of any existing force field. 
Its application requires the definition of an equilibrium structure where atoms are located in $\bRcal$, and the dielectric tensor $\boldsymbol{\epsilon}$ and the effective charges $Z_{i\alpha\beta}$ computed in this structure.
The equilibrium atomic positions $\bRcal$ should ideally correspond to a high symmetry phase, even if it is the saddle point of the energy landscape (with imaginary phonon frequencies). In this way, the energy and its derivatives satisfy all symmetry operations in subgroups of the parent one, thus enabling the potential to respect symmetry constraints of any subgroups. This is relevant in materials where temperature or pressure may change the crystallographic symmetry group by progressively increasing or decreasing the symmetries, like in perovskites, where the cubic Pm$\bar 3$m group is a parent of all other symmetry-broken phases.
The model's accuracy decreases as the structure deviates from the centroid one, particularly when the effective charges change significantly. However, in most polar crystals, effective charges are almost independent of the atomic positions as testified by the rarity of multi-phonon scattering modes observed in IR spectra experimentally\cite{monacelliTimedependentSelfconsistentHarmonic2021,Siciliano2023} originated by the dependency of effective charges on the atomic position. This assumption breaks down when simulating liquids, materials where ions diffuse\cite{grasselli_topological_2019}, or systems undergoing a significative rearrangement of chemical bonds (strongly first-order phase transitions). In the latter case, separate long-range models for each phase could be employed.

The only free parameter is the cutoff $\eta$: the distance below which the long-range interactions are smeared out. The lower $\eta$, the smaller the minimum distance between atoms interacting through dipole-dipole electrostatic, thus improving its accuracy. However, the computational cost to evaluate the energy and its derivative increases with low values of $\eta$ as we need to include more $k$-points in the summation of \eqname~\eqref{eq:energy:final}. Also, if $\eta$ is too small, higher orders of the multipole expansion, neglected by the polarization model, become important. Moreover, if the base short-range model has already been fitted, the long-range correction should not affect the training set. This sets a lower bound for $\eta$ to the maximum distance between any pair of atoms in the training data (accounting for PBC). Thus, the choice of $\eta$ should be a trade-off between computational cost, required accuracy, and the supercell dimension of the calculations in the training set. 
If $\eta$ is smaller than the maximum distance between atoms in the training data, the long-range model energies (\eqname~\ref{eq:energy:final}), forces (\eqname~\ref{eq:forces}), and stresses alter the training values.
In this case, the long-range contribution to the total energy (and derivatives), as evaluated by the model, must be removed from the training data, and the short-range force field must be retrained.

Once $\bRcal$, $Z_{i\alpha\beta}$, $\boldsymbol{\epsilon}$, and $\eta$ are established, the final total energy (and its derivatives) of any structure are obtained by adding the ones evaluated by the short-range force field with those of the model.

\subsection{Correcting phonon dispersions in \ch{BaTiO3}}

\begin{figure*}
    \centering
    \includegraphics[width=\textwidth]{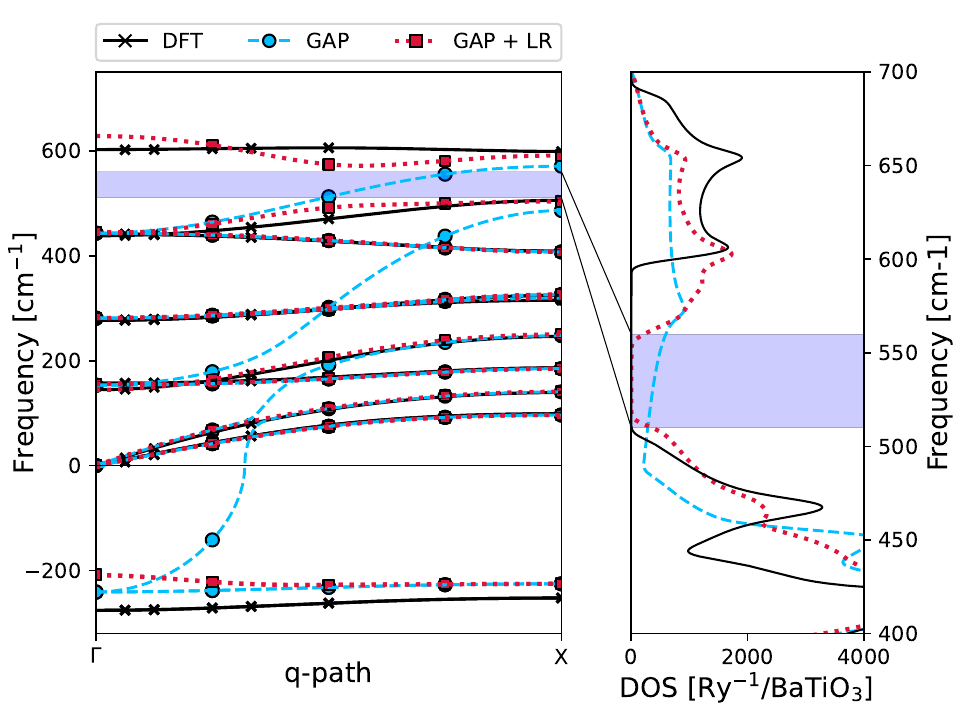}
    \caption{Phonon band structure of cubic \ch{BaTiO3}, comparison between DFT, the GAP without long-range (LR) interaction\cite{gigliThermodynamicsDielectricResponse2022}, and the same potential with long-range (LR) interactions adding to the forces \eqname~\eqref{eq:forces} Negative frequencies are imaginary numbers, meaning the structure is on a saddle point of the energy landscape. On the left, we report the dispersion along the $\Gamma-X$ high symmetry line. Scatter points are commensurate with the 8x8x8 supercell employed in the phonon calculation (DFPT calculations are reported with a black cross). The GAP without the long-range interactions fails to describe the phonon bands near $\Gamma$. The right panel is a zoom in the high-frequency phonon density of states (DOS), where the two bands above \SI{600}{\per\centi\meter} and below \SI{500}{\per\centi\meter} merge in the absence of long-range (LR) interactions. The value of the smearing parameter for the long-range (LR) interactions is $\eta = \SI{2.8}{\angstrom}$ }
    \label{fig:phonons}
\end{figure*}

Force engines that neglect long-range electrostatic interactions fail to recover the LO-TO splitting near the center of the Brillouin zone, resulting in a crude approximation of phonon dispersions.
This is one of the most severe sources of error in short-range force fields when applied to polar materials. 
We exemplify this issue in the instance of $\alpha$-\ch{BaTiO3} (cubic perovskite, five atoms in the primitive cell, crystallographic group 221, $Pm\bar 3 m$). Harmonic phonons of this system are characterized by two unstable (imaginary) harmonic phonons across the entire Brillouin zone, as this structure is stabilized by thermal fluctuations at high temperature\cite{gigliThermodynamicsDielectricResponse2022}. 
In \figurename~\ref{fig:phonons}, we report the phonons evaluated within a short-range Gaussian approximation potential (GAP) (Ref.~\cite{gigliThermodynamicsDielectricResponse2022}) with and without the addition of the long-range forces (\eqname~\ref{eq:forces}), with $\eta= \SI{2.5}{\angstrom}$. 
As a reference, we plot the first-principles phonon dispersions evaluated within DFPT (with the PBEsol exchange-correlation potential\cite{PBEsol}, more details on \appendixname~\ref{app:dft}), using the same parameters employed to generate the training set of the GAP model.
While the GAP already delivers a good agreement with DFPT at the edge of the Brillouin zone, it fails when $\bm q$ approaches $\Gamma$, displaying a significative deviation for some bands already at the halfway point of the Brillouin zone. This is 
also evident from the zoom-in of the density of states (DOS, right panel of \figurename~\ref{fig:phonons}), where a phonon band gap between \SI{500}{\per\centi\meter} and \SI{600}{\per\centi\meter} should separate the highest phonon band from the rest of the phonon dispersions. The short-range GAP does not recover this separation due to merging the LO and TO phonon bands near $\Gamma$. On the contrary, the long-range corrected force field qualitatively reproduces the phonon band gap, improving the agreement with DFPT. Another important deviation is observed in the cubic structure's imaginary (unstable) modes. In the short-range model, there is an additional imaginary mode nearby $\Gamma$, absent in both the long-range model and the DFPT dispersions.

The chosen value of $\eta$ (\SI{2.5}{\angstrom}) slightly modifies the phonon dispersion at the edge of the Brillouin zone, where the q points are commensurate with the 2x2x2 supercell employed in the training set of the GAP\cite{gigliThermodynamicsDielectricResponse2022}. If a larger supercell were employed in the training data, the long-range interactions would have significantly altered the phonon dispersion. In the latter case, the energy (and its derivatives) should have been pre-processed by removing the long-range contribution before the training.

%While this result was obtained without any retraining of the GAP, it is possible to further improve the models by subtracting the long-range energies (\eqname~\ref{eq:energy:final}) and forces (\eqname~\ref{eq:forces}) from the training set, thus improving the quality of the training as the remaining interactions decay in space as a the quadrupole-quadrupole, and are easier to fit with any method.

\section{Discussion}

This approach assumes that the dielectric tensor and effective charges are independent of atomic coordinates. 
This approximation fails when the change in the atomic position breaks covalent bonds or atoms can freely diffuse.
Thus, the method is more suited to describe the thermodynamics of structures that keep the same bonding network and topology during the simulation. However, it can still be used to study complex first-order phase transitions when coupled, e.g., with the SSCHA. Within this framework, atoms fluctuate around their centroids, and it is possible to compare the free energies of phases with different bonding networks without breaking any covalent bonds in the simulation, as demonstrated for the study of the $\gamma$-Y phase transition in metal-halides perovskites~\cite{Monacelli2023} or the hydrogen high-pressure phase-diagram~\cite{Monacelli2023_NatPhys}. 

Overcoming the limitations arising from the model assumptions requires
the introduction of the atomic environment dependency on effective charges, dielectric tensors, and centroids.
Electric properties like the total polarization, effective charges, and the dielectric tensor are, in fact, routinely parametrized through equivariant machine learning approaches\cite{Gastegger2017,Grisafi2021}, and the centroid parameter can be rigorously defined through the total polarization and effective charges by inverting \eqname~\eqref{eq:dipole:moment}. In this way, it is possible to replace in \eqname~\eqref{eq:energy:final} the explicit dependency of the effective charges, centroids, and dielectric tensor on the atomic positions, similarly as it has been proposed in third and fourth-generation machine learning force fields\cite{Behler2021}.

We presented a model that can pave the way to a rigorous introduction of long-range electrostatic interaction relying only on physical observables, i.e., without requiring extracting empirical parameters like local atomic charges. This energy-forces-stress model can be combined with existing short-range atomistic potentials, like the GAP shown in \figurename~\ref{fig:phonons}\cite{gigliThermodynamicsDielectricResponse2022},
to account for long-range electrostatic interactions and further improve their accuracy. The long-range part of energies and forces can be calculated efficiently within \eqname~\eqref{eq:energy:final} and \eqref{eq:forces}, respectively, and added to the results of short-range models. At the same time, the electrostatic stress tensor is computed by employing the algorithmic differentiation with the procedure explained in \eqname~(\ref{eq:stress:original},\ref{eq:R:strain},\ref{eq:k:strain}).
Only one free parameter needs to be tuned: the smearing $\eta$, quantifying the minimum range of the electrostatic interactions.
The Born effective charges $\bm Z$ and the dielectric tensor $\bepsel$ are evaluated from first-principles DFPT. The model can correctly account for LO-TO splitting and shines in condensed systems with fixed crystal structures, paving the way to the next-generation machine-learning force fields accounting for long-range electrostatic interactions.

\section*{Code avability}

The equations for energy and forces (\eqname~\ref{eq:energy}, \eqname~\ref{eq:forces}) have been implemented in Python and Julia\cite{JULIA} as an open-source package and can be downloaded from \href{https://github.com/mesonepigreco/electrostatic-calculator}{https://github.com/mesonepigreco/electrostatic-calculator}. The stress tensor is evaluated by exploiting the algorithmic differentiation of energy, as implemented in ForwardDiff.jl\cite{ForwardDiff}. The code is distributed under the GPLv3 license. It operates as a force field calculator for the Atomic Simulation Environment\cite{ASE} (ASE).

\section*{Data avability}
All the data reported in this work, such as dynamical matrices, dielectric tensor, and Born effective charges of \ch{BaTiO3} evaluated to produce \figurename~\ref{fig:phonons}, are published in the Example folder of the source code GitHub repository.

\section*{Acknowledgments}
L. M. acknowledges the H2020 program from the European Union for the MSCA-IF grant 101018714. This work was supported by the Swiss National Supercomputing Centre (CSCS) grant under project ID s1192.

    \appendix
    \section*{Appendices}
    \section{Auxiliary system of charges}
\label{app:charge:system}

We need to define a system of charges that satisfy the local dipole moment defined through \eqname~\eqref{eq:dipole:moment}. For this purpose, for each atom, we generate two charges of opposite sign and modulus $q_i$, ad a distance $d$ so that $\mu_i = q_i d$. The specific choice of $q_i$ and $d$ do not affect the electric field when $r\gg d$, so we have an arbitrary choice for their value. Without loss of generality, we define $q_i$ as the trace of effective charge of the atom $i$
\begin{equation}
q_i = \frac 13 \sum_\alpha Z_{i\alpha\alpha},
\label{eq:charge}
\end{equation}
and the position of the positive and negative charges $\tilde\bR_+$ and $\tilde\bR_-$ to align the center dipole moment in the middle of the atomic displacement:
\begin{subequations}
\begin{equation}
\tilde R_{i\alpha+} = R_{i\alpha} + \frac {1}{2q_i} \sum_\beta Z_{i\alpha\beta}(R_{i\beta} - \Rcal_{i\beta}) ,
\end{equation}
\begin{equation}
\tilde R_{i\alpha-} = R_{i\alpha} - \frac {1}{2q_i} \sum_\beta Z_{i\alpha\beta}(R_{i\beta} - \Rcal_{i\beta}) .
\end{equation}
\label{eq:rtilde}
\end{subequations}

    \section{Electric field}
\label{app:efourier}

The electric field generated by a charge distribution $\rho(\bm r)$ in a periodic system can be expressed more conveniently in Fourier space transforming the Maxwell relations:
\begin{equation}
    \nabla\cdot\bD = \rho,
\end{equation}
\begin{equation}
    \sum_{\alpha\beta} \ldielectric_{\alpha\beta} \frac{\partial E_\beta}{\partial r_\alpha} = \frac{\rho}{\varepsilon_0},
\end{equation}
\begin{equation}
    -\sum_{\alpha\beta} \ldielectric_{\alpha\beta} \frac{\partial^2 V}{\partial r_\alpha\partial r_\beta} = \frac{\rho}{\varepsilon_0},
\end{equation}
\begin{equation}
    -V(\bk)\sum_{\alpha\beta}\ldielectric_{\alpha\beta}k_\alpha k_\beta = \frac{\rho(\bk)}{\varepsilon_0},\label{eq:v:rho}
\end{equation}
where $\rho(q)$ is evaluated from the Fourier transform of \eqname~\eqref{eq:rho:r}
\begin{equation}
    \rho(\bk) = \frac{1}{\Omega}\int_\Omega d\br \sum_ie^{-i \bk\cdot\br} \frac{q_i}{\sqrt{8\pi^3} \eta^2} e^{-\frac{(\br - \bR_i)^2}{2\eta^2}}.
\end{equation}
Here, $\Omega$ is the volume of the simulation cell. $\bk$ can only assume values so that the integrated function is periodic in $\br$ (reported in \eqname~\ref{eq:k:vals}). %Note that we choose a gauge where we have a different phase factor for each term of the summation. This does not affect the result if we are consistent but strongly simplifies the calculation and makes the whole integrand a periodic function of $r$ 
Extending the integral over all the solid volume and dividing by the number of k-points $N_k$ used to sample the Brillouin zone, we can swap the integral and the summation
\begin{equation}
    \rho(\bk) = \frac{1}{N_k\Omega}\sum_ie^{-i\bk\cdot \bR_i}\int d\br e^{-i \bk\cdot(\br - \bR_i)} \frac{q_i}{\sqrt{8\pi^3} \eta^2} e^{-\frac{(\br - \bR_i)^2}{2\eta^2}},
\end{equation}
\begin{equation}
    \rho(\bk) = \frac{1}{N_k\Omega}\sum_ie^{-i\bk\cdot \bR_i}q_i e^{-\frac{\eta^2k^2}{2}}.
\end{equation}
Exploiting \eqname~\eqref{eq:v:rho}, one gets the electrostatic potential
\begin{equation}
    V(\bk) = -\frac{1}{N_k\Omega\varepsilon_0}\sum_i\frac{q_ie^{-i\bk\cdot \bR_i} e^{-\frac{\eta^2k^2}{2}}}{\sum_{\alpha\beta}k_\alpha \ldielectric_{\alpha\beta} k_\beta},
\end{equation}
and the electric field as
$$
\bE(\br) = -\nabla V(\br),\qquad
\bE(\bk) = -i\bk V(\bk):
$$
\begin{equation}
\bE(\bk) = \frac{i}{N_k\Omega}\sum_i \frac{q_i \bk e^{-i\bk\cdot \bR_i} e^{-\frac{\eta^2k^2}{2}}}{\sum_{\alpha\beta}k_\alpha \ldielectric_{\alpha\beta} k_\beta}.
\end{equation}
To return to real space, we perform the inverse Fourier transform summing over all the $k$ points, excluding the $\Gamma$ ($\bk = 0$), as that term is zero in neutral systems ($\sum_i q_i = 0$):
\begin{equation}
    \bE(\br) = \frac{i}{N_k\Omega}\sum_{\substack{ij\\ k_j\neq 0}} \frac{q_i \bk_j e^{-i\bk_j \cdot \bR_i}e^{-\frac{\eta^2k_j^2}{2}}}{\sum_{\alpha\beta}{k_j}_\alpha \ldielectric_{\alpha\beta} {k_j}_\beta}e^{i\bk_j\cdot \br},
\end{equation}
\begin{equation}
    \bE(\br) = \frac{i}{N_k\Omega}\sum_{\substack{ij\\ k_j\neq 0}} \frac{q_i \bk_j e^{-\frac{\eta^2k_j^2}{2}}}{\sum_{\alpha\beta}{k_j}_\alpha \ldielectric_{\alpha\beta} {k_j}_\beta}e^{i\bk_j\cdot (\br - \bR_i)}.
\end{equation}
The sum over $i$ goes on all atoms in all the cells, while $j$ only runs on the cells (the k-points), and the order of the two series can be exchanged thanks to the exclusion of $k_j = 0$.
For convenience, we define the charge structure factor as
\begin{equation}
    S(\bk) = \frac{1}{N_k}\sum_i q_i e^{-i \bk \cdot \bR_i}.
    \label{eq:Sk:old}
\end{equation}
Since $e^{-i \bk \cdot \bR_i}$ is invariant if we shift the atomic positions $\bR_i$ by a lattice vector, the summation on the atoms $i$ in the supercell is $N_k$ times the one evaluated in the primitive cell only (\eqname~\eqref{eq:Sk}).
Thus, we get the expression of the electric field
\begin{equation}
    \bE(\br) = \frac{i}{\Omega}\sum_{\substack{j\\ k_j\neq 0}} \frac{\bk_j e^{-\frac{\eta^2k_j^2}{2}}e^{i\bk_j\cdot \br}}{\sum_{\alpha\beta}{k_j}_\alpha \ldielectric_{\alpha\beta} {k_j}_\beta}S(\bk_j),
\end{equation}
that coincides with the one reported in the main text (\eqname~\ref{eq:E:fourier}).

\section{Expression of the energy}
\label{app:energy}
Thanks to \eqname~\eqref{eq:charge} and \eqref{eq:rtilde}, it is possible to build a system of point charges whose positions depend on the atomic coordinates $\bR$ and the equilibrium configuration $\bRcal$.

Substituting the expression of the dipoles (\eqname~\ref{eq:dipole:moment}) in the energy \eqname~\eqref{eq:energy}, we get:
\begin{equation}
\mathcal{E}= -\frac 12\sum_{i\alpha\beta} (R_{i\alpha} - \Rcal_{i\alpha}) Z_{i\beta\alpha} E_{\beta}\left(\bR_i\right).
\label{eq:energy2}
\end{equation}
where $E_\beta$ is the $\beta$ Cartesian component of the electric field on the atom. Substituting the expression of the electric field obtained from the Ewald sum (\eqname~\ref{eq:E:fourier}) into \eqname~\eqref{eq:energy2}, we get
\begin{align}
    \mathcal E = &- \frac 12\sum_{i\alpha\beta}(R_{i\alpha} - \Rcal_{i\alpha})\frac{i}{\Omega}\sum_{\substack{k,j\\k\neq 0}}\frac{Z_{i\beta\alpha}k_\beta e^{-\frac{\eta^2k^2}{2}}}{\sum_{\mu\nu} k_\mu\ldielectric_{\mu\nu}k_\nu} q_je^{-i\bk(\bR_j - \bR_i)}\cdot \nonumber \\
    &  \cdot\left[
    e^{-i\sum_{\mu\nu} \frac{k_\nu Z_{j\nu\mu}}{2q_j} (R_{j\mu} - \Rcal_{j\mu})} - 
    e^{i\sum_{\mu\nu} \frac{k_\nu Z_{j\nu\mu}}{2q_j} (R_{j\mu} - \Rcal_{j\mu})}\right].
\end{align}
\begin{align}
    \mathcal E = \frac 12\sum_{i\alpha\beta}&(R_{i\alpha} - \Rcal_{i\alpha})\frac{1}{\Omega}\sum_{\substack{k,j\\k\neq 0}}\frac{Z_{i\beta\alpha}k_\beta e^{-\frac{\eta^2k^2}{2}}}{\sum_{\mu\nu} k_\mu\ldielectric_{\mu\nu}k_\nu} \cdot \nonumber \\
    & \cdot2q_j e^{-i\bk(\bR_j - \bR_i)}\sin\left(\sum_{\mu\nu} \frac{k_\nu Z_{j\nu\mu}}{2q_j} (R_{j\mu} - \Rcal_{j\mu})\right)
\end{align}
Since $k$ is small (constrained by the exponential $e^{-\eta^2 k^2/2}$, where $\eta$ is large), we expand the $\sin$ up to the first order and get \eqname~\eqref{eq:energy:final}:
\begin{align}
    \mathcal E = \frac 12\sum_{ij\alpha\beta\mu\nu}&(R_{i\alpha} - \Rcal_{i\alpha})(R_{j\mu} - \Rcal_{j\mu})\frac{Z_{i\beta\alpha}Z_{j\nu\mu}}{\Omega} \cdot \nonumber \\ &\cdot\sum_{\substack{k\\k\neq 0}}\frac{k_\beta k_\nu e^{-\frac{\eta^2k^2}{2}}}{\sum_{\mu\nu} k_\mu\ldielectric_{\mu\nu}k_\nu} e^{-i\bk(\bR_j - \bR_i)} .
\end{align}
As pointed out in the main text, the dependency from $q$ disappears when expanding the sinus up to the first order, removing any arbitrary choice of the charged system to represent the correct polarization.

    \section{Acoustic sum rule}

\label{app:asr}
The forces equation (\eqname~\ref{eq:forces}) violates the translational invariance. 
While the acoustic sum rule in the effective charge prevents a rigid translation of the system from generating a finite dipole moment
\begin{equation}
    \sum_i Z_{i\alpha\beta} = 0\qquad \forall \alpha, \beta;
\end{equation} 
it cannot prevent a global translation from affecting the atomic local dipole $\bm\mu_i$, resulting in a shift of energy and a nonzero net force on the center of mass.
To prevent this dependence of the local atomic dipole on global translations, we redefine the centroid position $\bRcal$ to eliminate rigid shifts
\begin{equation}
    \Rcal_{i\alpha} = 
\Rcal_{i\alpha}^{(0)}
 + 
\frac 1 N \sum_{j} (R_{j\alpha} - \Rcal_{j\alpha}^{(0)}).
\label{eq:rcal:asr}
\end{equation}
With this redefinition, the overall energy function (\eqname~\ref{eq:energy:final}) is invariant under global translations. Thus, its derivatives satisfy all the acoustic sum rules.
In particular, we show that the actual contribution of this redefinition of the centroid on the forces cancels the force on the center of mass present in \eqname~\eqref{eq:forces}:
\begin{equation}
    f^{\text{ASR}}_{i\alpha} = -\sum_{j\beta}\frac{\partial \mathcal E}{\partial \Rcal_{j\beta}}\frac{\delta_{\beta\alpha}}{N} = \frac 1N \sum_j\frac{\partial \mathcal E}{\partial R_{j\alpha}} = -\frac 1 N \sum_j f_{j\alpha}.
    \label{eq:f:asr}
\end{equation}
In the numerical implementation, this expression can be applied \emph{a posteriori} to \eqname~\eqref{eq:forces}.

Care must be taken also when computing the stress tensor. In our implementation, since the redefinition of $\bRcal(\bR)$ (\eqname~\ref{eq:rcal:asr}) occurs inside the function that computes the total energy (\eqname~\ref{eq:energy:final}), the chain rule performed by the algorithmic differentiation automatically includes also the derivatives of the $\bRcal[\bR(\bm \varepsilon)]$, thus correcting the ASR also for the stress.
    \section{Details of the DFT calculation}
\label{app:dft}

For the calculation of the DFT phonon spectrum of \ch{BaTiO3} cubic perovskite reported in \figurename~\ref{fig:phonons}, as well as for evaluating the effective charges and dielectric tensor, we employed the suite Quantum ESPRESSO\cite{Giannozzi2017} version 7.0 within the PBEsol exchange-correlation potential approximation\cite{PBEsol}.
We used PAW and ultrasoft pseudopotentials from the SSSP library version 1.2.1-efficiency\cite{SSSP}, with a cutoff for the wavefunction and density of \SI{60}{\rydberg} and \SI{600}{\rydberg}, respectively.
We sampled the Brillouin zone for the electrons with an 8x8x8 k-mesh without offset.
The lattice parameter used for the simulation is \SI{4.035}{\angstrom}. 
    %\input{src/stress.tex}
    %\input{src/appendix.tex}

    %\bibliographystyle{ieeetr}
	%\bibliography{biblio}
	%\input{main.bbl}

\end{document}